\documentclass[10pt,conference]{IEEEtran}

\usepackage{physics}
\usepackage{bbold}
\usepackage{amsthm}
\usepackage{amsmath}
\usepackage{graphicx}
\usepackage{subfigure}
\usepackage[linesnumbered,ruled,vlined]{algorithm2e}

\newcommand\numberthis{\addtocounter{equation}{1}\tag{\theequation}}

\title{The Impact of Logical Errors on Quantum Algorithms}

\author{\IEEEauthorblockN{Omer Subasi}
\IEEEauthorblockA{\textit{Pacific Northwest National Laboratory (PNNL)}\\
Richland, Washington, USA \\
omer.subasi@pnnl.gov}
\and
\IEEEauthorblockN{Sriram Krishnamoorthy}
\IEEEauthorblockA{\textit{Google}, USA \\
sriram.krishnamoorthy@gmail.com
}
}

\date{
}

\begin{document}

\maketitle

\pagestyle{plain}

\begin{abstract}
In this work, we explore the impact of logical stochastic Pauli and coherent Z-rotation errors on quantum algorithms. We evaluate six canonical quantum algorithms' intrinsic resilience to the logical qubit and gate errors by performing the Monte Carlo simulations guided by the quantum jump formalism. The results suggest that the resilience of the studied quantum algorithms decreases as the number of qubits and the depth of the algorithms' circuits increase for both Pauli and Z-rotation errors. Our results also suggest that the algorithms split into two different groups in terms of algorithmic resilience. The evolution of Hamiltonian, Simon and the quantum phase estimation algorithms are less resilient to logical errors than Grover's search, Deutsch-Jozsa and Bernstein-Vazirani algorithms.
\end{abstract}


\section{Introduction}
Current quantum computing devices are experiencing quantum (noise) errors that inhibit the success of quantum algorithms \cite{bharti2022noisy, Leymann_2020}. 
Quantum errors are the result of the interactions in open quantum systems.
These errors can be either stochastic or coherent \cite{devitt2013quantum, failuredist, Greenbaum_2017}. Pauli errors are well-known stochastic errors. On the other hand, coherent errors are systematic (rotation) errors such as Z-rotation errors around the Z axis. While there has been wide-ranging research on understanding the impact of these errors on variational algorithms \cite{ding2022evaluating, fontana2021evaluating, quiroz2021quantifying}, specific quantum circuits \cite{Reiner_2018, xue2019effects}, quantum compiling \cite{sharma2020noise}, or with limited scope \cite{berberich2023robustness, volya2020special}, the impact \emph{logical} stochastic and coherent errors on canonical quantum algorithms has not been studied in the same manner. 
Therefore, in this study, we evaluate the impact of logical stochastic and coherent errors on six quantum algorithms exploring their algorithmic resilience. In particular, we explore \emph{gate and qubit logical errors} which are the abstraction that an algorithm faces as opposed to physical noise such as those caused by quantum decoherence, circuit artifacts, and cross interactions. Studying logical errors provides a direct way to assess intrinsic algorithmic resilience to quantum errors. 

Our timely work is setting a beginning in understanding the effects of logical errors whose current occurrence rates ($1.7 \times 10^{-6}$ per cycle) can now allow the logical performance scaling with increasing qubit numbers as reported in a recently announced breakthrough by Google Quantum AI \cite{google2023suppressing} where a logical qubit prototype was successfully built. 

The framework we develop is based on state-vector simulations that are guided by the \emph{quantum jump formalism (method)} \cite{PhysRevLett.68.580, Molmer:93}. The quantum jump method is equivalent to the Lindblad master equation in describing the evolution of open quantum systems represented by a density matrix. However, it has significant computational advantages. The quantum jump method based simulations have $O(2^n)$ time and memory complexity while Lindbladian simulations have $O(2^{2n})$ time and memory complexity for $n$ qubits. Because of these reasons, in this work, we perform state-vector simulations based on the quantum jump formalism to assess the impact of logical quantum errors on quantum algorithms. In summary,
\begin{itemize}
    \item We evaluate the impact of logical Pauli and coherent Z-rotation gate and qubit errors on six quantum algorithms.
    \item To perform our evaluations, we implement a software framework with an error injection routine that modifies an algorithm's Directed Acyclic Graph (DAG) in Qiskit \cite{Qiskit}. 
    \item The results suggest that the resilience of the studied quantum algorithms decreases as the number of qubits and the depth of the algorithms' circuits increase for both Pauli and Z-rotation errors.
    \item Considering intrinsic algorithmic resilience, the algorithms form two different groups. The algorithms of the evolution of Hamiltonian, Simon and the quantum phase estimation are less resilient than Grover's search, Deutsch-Jozsa and Bernstein-Vazirani algorithms. 
\end{itemize}

The organization of this paper is as follows:
Section \ref{background} provides the background for the formalisms of quantum systems.
Section \ref{monte} formulates the number Monte Carlo simulations of open quantum systems for a desired accuracy.
Section \ref{frame} presents our software framework.
Section \ref{eval} discusses the experimental evaluation and the results.  
Section \ref{related} overviews the related work.
Section \ref{conclusion} summarizes this study.

\section{Background: Quantum Systems Formalisms}
\label{background}
In this section, we first introduce the mathematical formalisms describing closed and open quantum systems. We then discuss the equivalent Lindbladian and quantum jump formalisms that describe the evolution of open quantum systems.

\subsection{Closed and Open Quantum Systems}
A quantum system can be either a closed or an open system.
Closed systems are ideal systems that do not interact with the environment.
As a result, they do not experience quantum errors.
In contrast, in open quantum systems, the system interacts with the environment and thus experiences errors.

Closed quantum systems are formalized based on the state-vector of pure states. The time evolution of a closed system is described by an unitary operator $U$ on an initial state $\psi(t_0)$ such that
\begin{equation}
\psi(t)  =  U(t)\psi(t_0).
\end{equation}

Realistically almost all quantum systems interact with their environments due to \emph{quantum decoherence}.  To characterize open quantum systems, a formalism called \emph{density matrix (operator)}, is often used. 
The density matrix describes the mixed (incoherent) quantum states.

The density matrix $\rho$ for the pure state $\ket{\psi}$ is given by $\rho := \dyad{\psi}{\psi}$.
The density matrix for statistical mixtures of quantum states $\ket{\psi_i}$ is
\[\rho^{mix} = \sum_i p_i \dyad{\psi_i}{\psi_i}\]
where $ \sum_i p_i = 1$.

The evolution of an open system $S$ and its environment $E$  is described in the total product Hilbert space $H_S \otimes H_E$.
Assuming the initial state of the combined density matrix $\rho = \rho_S \otimes \dyad{0}{0}_E$, the evolution of
the total system is described by
\begin{equation}
\rho(t) = U_{SE}  (\rho_S \otimes \dyad{0}{0}_E)  U^{\dagger}_{SE}.
\end{equation}

The evolution of the system $S$ is obtained by a partial trace on $E$ 
\begin{align*}
\rho_S(t) &  = \Tr_E{[\rho_S(t)]}\\ 
&= \sum_i \expval{U_{SE}  (\rho_S \otimes \dyad{0}{0}_E)  U^{\dagger}_{SE}}{i} \\
& = \sum_i \expval{U_{SE} \ket{0} \rho_S(0) \bra{0}  U^{\dagger}_{SE}}{i}, \numberthis
\end{align*}
where $\ket{i}$ is some orthonormal basis for the environment. 



\subsection{The Lindblad Master Equation}
The Lindblad master equation is the most general type of differential Markovian master equation describing 
the generally non-unitary evolution of a density matrix $\rho$. 
In the most general form, for a system density matrix $\rho_S$ 
\begin{equation} \label{lindblad}
\dv{\rho_S}{t}  =  -\frac{i}{\hbar}\comm{H_S}{\rho_S} + \pounds(\rho_S)
\end{equation}
where $H_S$ is the Hamiltonian of the system and $\comm{.}{.}$ is the commutator operator and $\pounds$ is a super-operator
characterizing the interactions between the system and the environment. This super-operator has the form 
\begin{equation} \label{jumpops}
\pounds(\rho_S)  =  \sum_i \gamma_i  J_i \rho_S J^{\dagger}_i - 
\frac{1}{2} \sum_i \gamma_i ( J^{\dagger}_i J_i \rho_S +  \rho_S  J^{\dagger}_i J_i)
\end{equation}
where $J_i$ is a \emph{jump operator} describing the interactions and $\gamma_i$ is some constant.

\subsection{Quantum Jump Method}
The \emph{quantum jump method} was introduced by \cite{PhysRevLett.68.580, Molmer:93} 
within the quantum optics research. 
This jump method is a stochastic Monte Carlo procedure describing the time evolution of
an open quantum system. The jump method describes independent realizations of the time evolution
starting with an initial state $\ket{\psi (t)}$.

To see how the jump method works, we proceed as follows:
First, to derive the overall non-hermitian Hamiltonian $H$, we write Equation \ref{lindblad} explicitly,
\begin{align*} \label{explicitlinblad}
\dv{\rho_S}{t}  =  
-\frac{i}{\hbar} \bigg[ \Big(H_S - \frac{i}{2} \sum_i \gamma_i  J^{\dagger}_i J_i\Big) \rho_S - \\ \rho_S \Big(H_s+\frac{1}{2} \sum_i \gamma_i  J^{\dagger}_i J_i\Big) \bigg] \\ 
+ \sum_i \gamma_i  J_i \rho_S J^{\dagger}_i \\ 
 = -\frac{i}{\hbar}  H \rho_S - \rho_S H^{\dagger} + \sum_i \gamma_i  J_i \rho_S J^{\dagger}_i \numberthis 
 \end{align*} 
 and we see that
\begin{equation} \label{simlinblad}
H = H_S - \frac{i \hbar}{2} \sum_i \gamma_i  J^{\dagger}_i J_i.  
\end{equation}
In Equation \ref{simlinblad}, the first term is interpreted as Schrodinger's evolution of pure states 
and the last term is interpreted with the
\emph{quantum jump} operators $J_i$ such that $\rho_S $ evolves into $J_i(\rho_S)$ with some probability.

Taking the Hamiltonian $H$ in Equation \ref{simlinblad},  
the stochastic time evolution for a quantum state $\ket{\psi (t)}$  is given by
\begin{equation}
\dv{\ket{\psi (t)}}{t} = (1 - \frac{i \delta t}{\hbar}  H) \ket{\psi (t)}.
\end{equation}
After infinitesimal time $\delta t$, if no \emph{jumps} occurs, then the state evolves into
\begin{equation}\label{nojump}
\ket{\psi(t+\delta t)} =  \frac{(1 - \frac{i \delta t}{\hbar}  H) \ket{\psi(t)}}{\sqrt{1-\delta p}}
\end{equation}
where
\begin{equation}
\delta p =  \sum_i \delta p_{i}  = \delta t \sum_i \gamma_i \expval{J^{\dagger}_i J_i}{\psi}
\end{equation}
and a normalization is introduced since the Hamiltonian $H$ in not hermitian.
If a jump $J_i$ occurs, then the state evolves into
\begin{equation}\label{jump}
\ket{\psi (t+\delta t)} = \sqrt{\frac{\gamma_i \delta t}{\delta p_{i} }} J_i \ket{\psi(t)}.
\end{equation}
The overall interpretation is that the system evolves into two possible outcomes at time $t$:
\begin{enumerate}
\item with probability $1 - \delta p$, it evolves into the state $\ket{\psi(t + \delta t)}$ as described by Equation \ref{nojump} and
\item with probability $\delta p_i$, it jumps to another state. The possible states are described Equation \ref{jump} by the jump operators $J_i$ with probability $\delta p_{i}$.
\end{enumerate}

The quantum jump method is general and applicable to many quantum-mechanical systems.
Here, we employ the jump method for error simulation in quantum algorithms. 
In this application of the jump method, error operators correspond to the jump operators. 
That is, the error gates, such as X, Y and Z gates, represent the jump operators in an algorithm's circuit. 
Since we simulate logical errors, at the time of an error injection, a jump described by the error gate is inserted.
The algorithm continues from the state the error gate led to.

\section{Monte Carlo Simulations in the Jump Method}
\label{monte}
An important question in a Monte Carlo simulation is to choose the number of independent
runs so as to obtain an accurate estimation.
With regard to the number of independent runs in the quantum jump method, it affects not only the accuracy of the
estimation but also the relative computational complexity to the density matrix simulation based on the Lindblad master
equation. 

Given $n$ qubits, the size of a state-vector of a quantum system is $N = 2^n$ in the jump method 
and $N^2 = 2^{2n}$ in a density matrix simulation. 
While a single run of the density matrix method would be enough to obtain the probabilities in the outcomes, 
the jump method requires independent runs to obtain accurate estimates of the outcomes.

Our goal is to find a function $f(\sigma)$ that gives the number of runs achieving 
the desired accuracy $\sigma$ for a Monte Carlo error simulation.
Following~\cite{BREUER199746}, we want to find a general function 
which does not depend on the specifics of a simulation method
but rather depends on the observables of the system of interest and the desired accuracy. 

Let $O$ be an observable of interest of an open quantum system. 
Let $\sigma_O$ be the standard error of the observable $O$.
We can use the sample average as an estimator $\hat{O_t}$ for the expected value of $O$ at time $t$:
\begin{equation}
\hat{O_t} = \frac{1}{f(\sigma_O)} \sum_{i=1}^{f(\sigma_O)} \ev{O}{\psi^i(t)}
\end{equation} 
where $\psi^i$ is the the quantum state of the system at $i$-th run of the simulation. 
The statistical error can be estimated by
the square root of the variance of $\hat{O_t}$
\begin{equation}
\sigma_O := \sqrt{Var(\hat{O_t})}. 
\end{equation}
The variance of $\hat{O_t}$ at time $t$ is
 \begin{equation}
Var(\hat{O_t}) = \frac{1}{f(\sigma_O)} Var(\ev{O}{\psi^i(t)}).
\end{equation} 
Therefore, the standard error of the mean can be estimated by
\begin{equation}
\sigma^2_O = \frac{\lambda_O(N)}{f(\sigma_O)}
\end{equation}
where $\lambda_O(N)$ is a factor quantifying the variance of the observable $O$ as a function of $N$. 
We note that this factor does not depend on the number of runs $f(\sigma_O)$.
From the last equation, we find $f(\sigma_O)$ as
\begin{equation}
f(\sigma_O) = \frac{\lambda_O(N)}{\sigma^2_O}.
\end{equation}

As investigated in \cite{PhysRevA.98.022111},  the time evolution of quantum systems is a
\emph{self-averaging} process which means the variance of the observables scales with $N^{-1}$. 
As a result, we can assume $\lambda_O(N) \propto N^{-1}$.
Therefore, we finalize $f(\sigma_O)$ as
\begin{equation}\label{numsim}
f(\sigma_O) = \frac{1}{\sigma^2_O N}
\end{equation}
where $\sigma_O$ is the desired standard error of the observable $O$.

The jump method simulating the state-vectors is not only significantly less expensive than 
the density matrix method but also is easily parallelizable. 
This is because each independent run of the jump method can be run in parallel with others.
However, the parallelization of density matrix simulations is highly non-trivial.

\section{The Software Framework}
\label{frame}
We design and implement an error analysis software framework for quantum algorithms within Qiskit \cite{Qiskit}.
Qiskit provides a comprehensive framework that includes the implementation of many quantum algorithms and 
tools to produce, manipulate, and simulate algorithm circuits and DAGs.

Our error model targets \emph{logical quantum errors}. It includes both logical qubit and gate errors.
Our model assumes errors occur at an algorithmic level where either i)
an algorithm's gate (operation) may not produce the correct result or ii) a logical qubit experiences an error such that its state is no longer coherent and deviates from the intended state. 
In effect, we abstract away the errors occurring 
at the physical quantum machine or circuit, and assume that they manifest as a flawed logical gate or a corrupted logical qubit. 
Our model includes stochastic Pauli and Z-rotation errors.
While Pauli X, Y, and Z errors are probabilistic in nature, Z-rotation errors are deviations of quantum operations 
from the ideal ones with a certain angle around the Z-axis of the Bloch sphere.
We only study Z-rotations as a representative of coherent rotations. 
   
Our software framework modifies the algorithms' DAG circuits. 
An error gate is appended to the specific logical gate or operation that is selected to be flawed in the DAG circuit. Similarly, an error gate is applied to a logical qubit that is selected to have a corrupt state.
Then, the correct and the corrupt DAGs are executed and their results are compared.
Algorithm \ref{algerror} shows our error injection routine.
\begin{algorithm}[!ht]
\DontPrintSemicolon
  \KwIn{DAG $dag$, error gate $error$, error index $index$}
  \KwOut{Modified DAG $dag$}
  \KwData{Add $error$ to $dag$}
  opnode = $dag$.op\_nodes()[$index$]\\
  mini\_dag = DAGCircuit()\\
  mini\_dag.\_add\_op\_node(opnode)\\
  p = QuantumRegister(len(opnode.qargs), "p")\\
  mini\_dag.add\_qreg(p)\\
   \For{$pq \in p$}{
		mini\_dag.apply\_operation\_back($error$, qargs=[pq])\\
	}
	wires = p\\
    $dag$.substitute\_node\_with\_dag(node=opnode, input\_dag=mini\_dag, wires=wires)         
\caption{Error Injection Routine}
\label{algerror}
\end{algorithm}

\begin{figure}[!ht]
\centering
     \begin{subfigure}
     \centering
\includegraphics[scale=0.5]{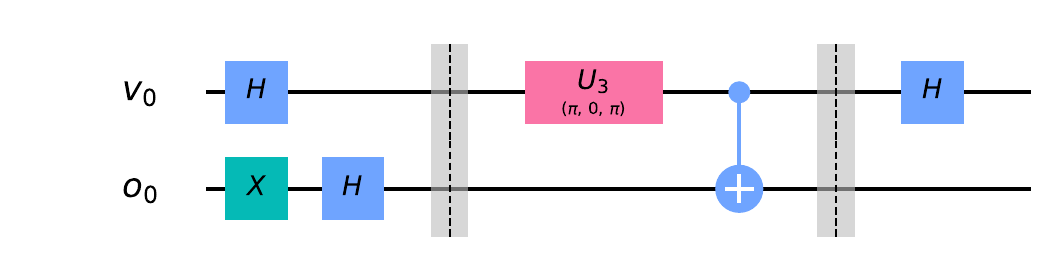}
 \end{subfigure}  
     \begin{subfigure}
     \centering
\includegraphics[scale=0.5]{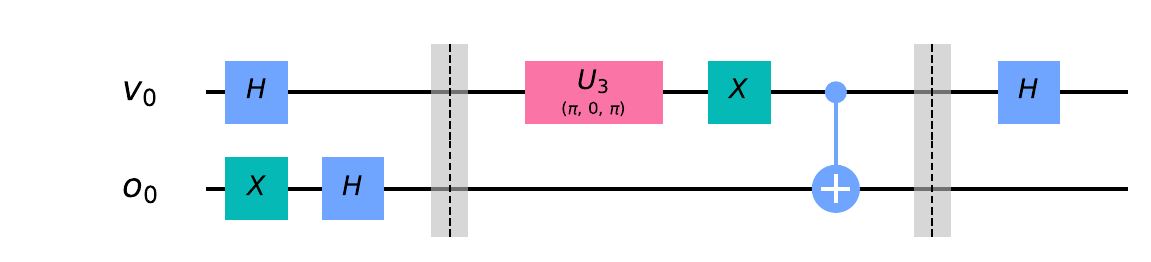} 
    \end{subfigure} 
           \caption{The original and the Pauli X error-injected circuits.}
            \label{fig:circs}
\centering
     \begin{subfigure}
     \centering
\includegraphics[scale=0.6]{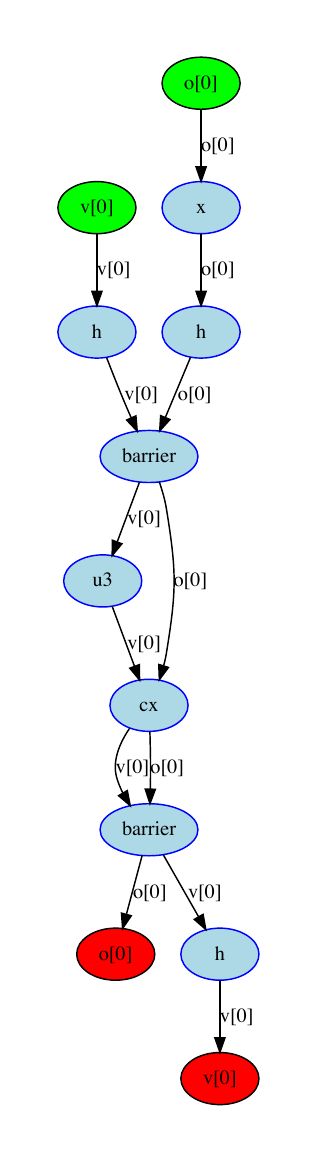} 
     \end{subfigure}
     \quad 
      \begin{subfigure}
      \centering
\includegraphics[scale=0.6]{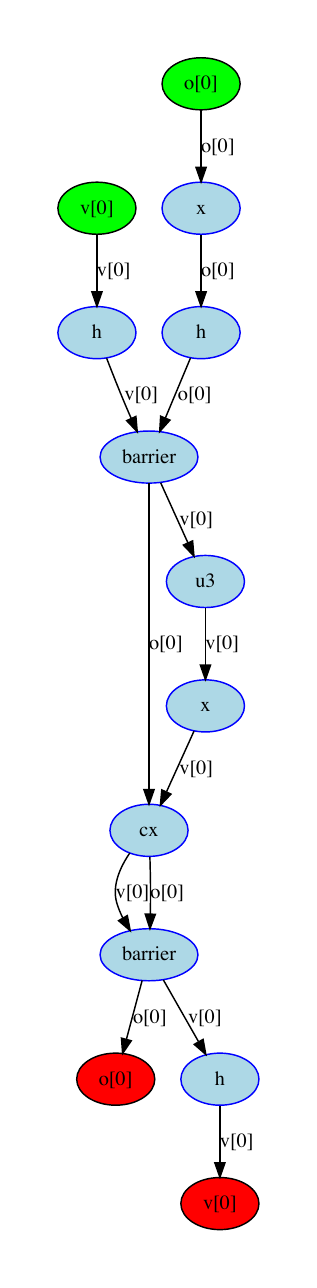}
     \end{subfigure} 
           \caption{The original and the Pauli X error-injected DAGs.}       
 \label{fig:dags}
\end{figure}

As an example, Figure \ref{fig:circs} shows a Bernstein-Vazirani circuit and its modified variant where a Pauli X
error injected. Figure \ref{fig:dags} shows the two corresponding DAGs.

\begin{figure*}[!ht]
\centering
     \begin{subfigure}
         \centering
\includegraphics[width=0.32\textwidth]{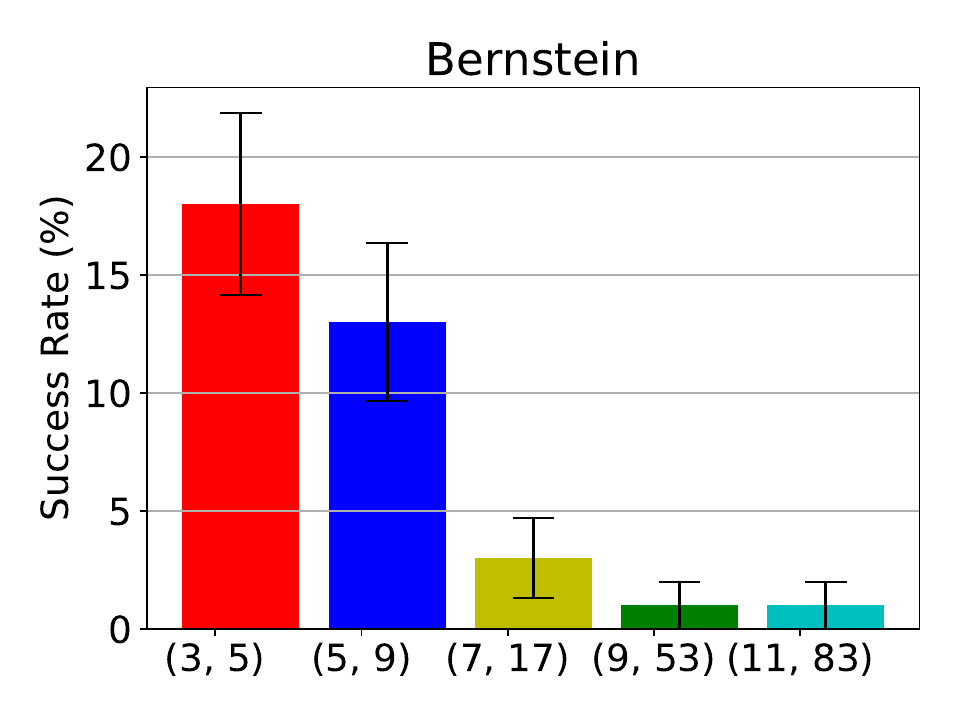} 
         \label{fig:bern}
     \end{subfigure}
     \begin{subfigure}
         \centering
\includegraphics[width=0.32\textwidth]{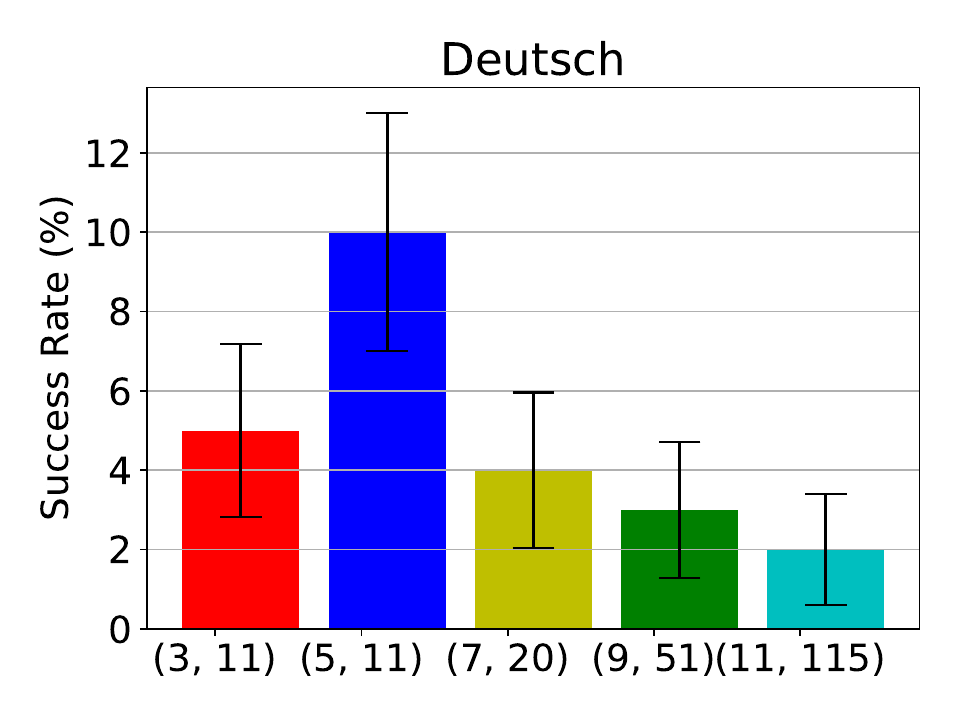} 
         \label{fig:deutsch}
     \end{subfigure}   
     \begin{subfigure}
         \centering
\includegraphics[width=0.32\textwidth]{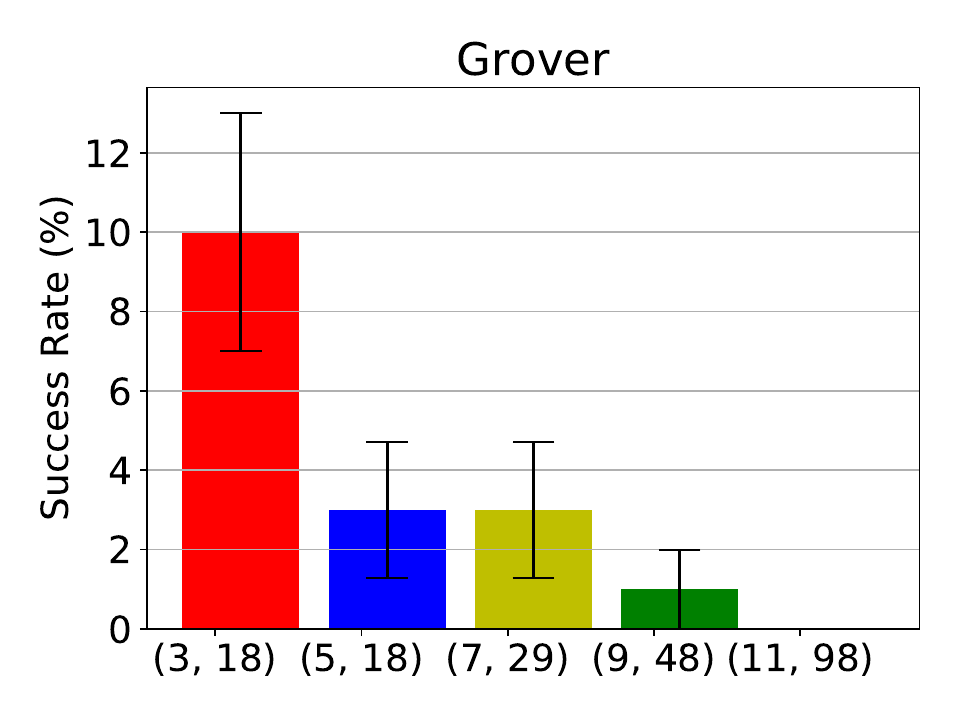} 
         \label{fig:grove}
     \end{subfigure}
     \begin{subfigure}
         \centering
\includegraphics[width=0.32\textwidth]{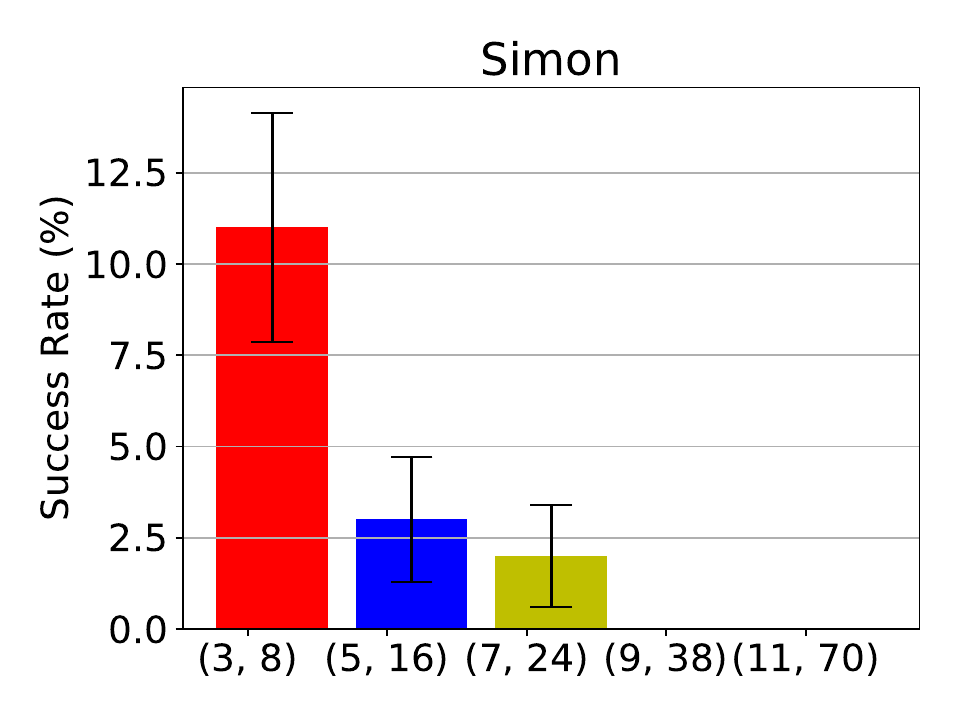} 
         \label{fig:simon}
     \end{subfigure}
     \begin{subfigure}
         \centering
\includegraphics[width=0.32\textwidth]{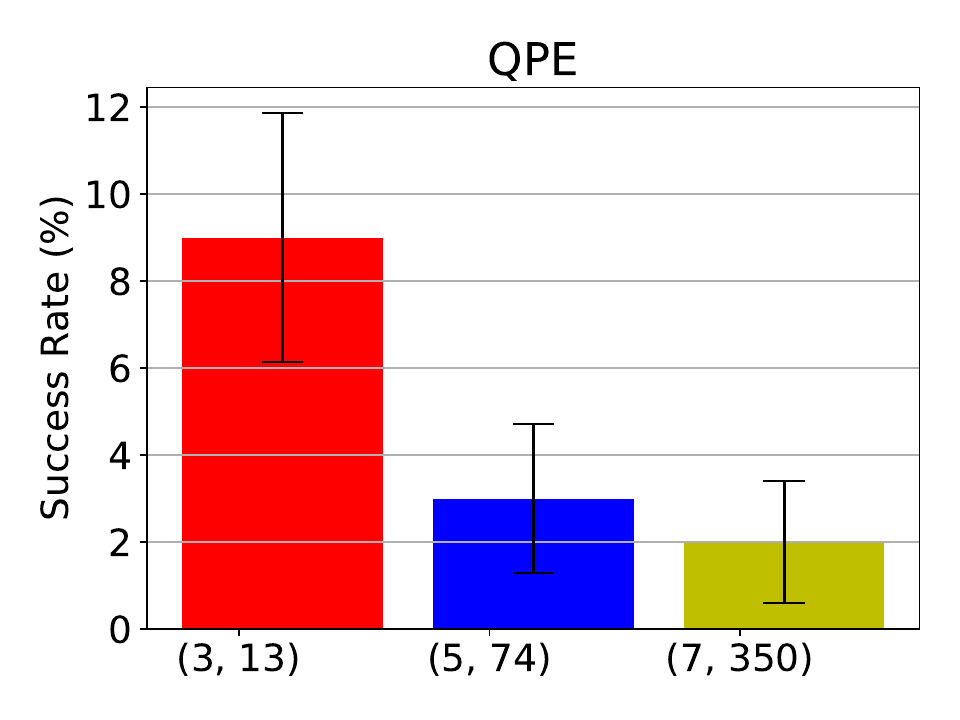} 
         \label{fig:qpe}
     \end{subfigure}
     \begin{subfigure}
         \centering
\includegraphics[width=0.32\textwidth]{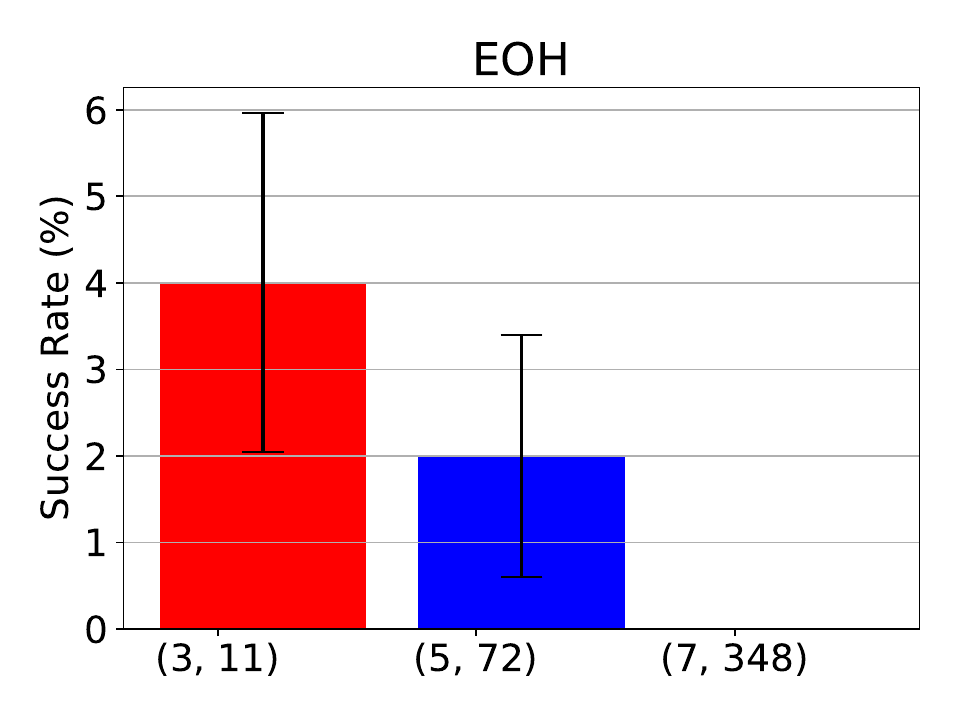} 
         \label{fig:eoh}
     \end{subfigure}
      \caption{Single Pauli Errors Results (\%)}
        \label{fig:paulires}
\end{figure*}

\section{Experimental Evaluation}
\label{eval}
In this section, we first present the experimental setup and then discuss the results.
\subsection{Experimental Setup}
We perform Monte Carlo simulations to evaluate
the impact of logical errors for six quantum algorithms in Qiskit \cite{Qiskit} (version 0.19.3) and Python 3.8.
Our simulations were run on a single node of Koothan cluster at PNNL. A single Koothan node has an Intel Xeon CPU running at 2.20GHz with 28 cores and 8 GBs of memory. 
Our performance metric is \emph{success ratio (percentage)}:
It is defined as the ratio of the successful runs - in which an algorithm
produces the ideal result - to the total number of runs.
The total number of runs is determined by Equation \ref{numsim} where we set
the target standard error to $\sigma_O = 0.01 \  (1\%)$. Let $N = 2^n$ where $n$ is the number of
qubits. We conservatively set $N=2^3=8$ in all simulations to increase the confidence in the results. 
Consequently, we run each case $\frac{1}{8 \times 10^{-4}} = 1250$ times.

Pauli errors are injected with X, Y, and Z gates. Z-rotation errors are injected with RZ gate that performs a rotation
around the Z-axis with the given angle. The evaluated angles are 
$\frac{\pi}{2}$, $\frac{\pi}{4}$, $\frac{\pi}{8}$, $\frac{\pi}{16}$ and $\frac{\pi}{32}$. Moreover, 
only single and double errors are evaluated. 
Because the algorithms always fail with double errors, higher numbers of errors are not tested.
The place of an error injection, i.e., the specific corrupt circuit gate or qubit, is selected uniformly at random.

We evaluate the following algorithms:

The Deutsch-Jozsa algorithm \cite{deutsch1992rapid} is one of the first quantum algorithms that showed an
exponential speedup compared to a deterministic classical algorithm,
given a blackbox oracle function. The algorithm determines whether the given function
$f:\{0,1\}^n \rightarrow \{0,1\}$ is constant or balanced. A constant function maps all 
inputs to 0 or 1, while a balanced function maps half of its inputs to 0 and the other half to 1.

The Bernstein-Vazirani algorithm \cite{bernstein}  determines a secret string $s \in \{0,1\}^n$ that maps $f:\{0,1\}^n \rightarrow \{0,1\}$ such that
$f(x)=s \cdot x \  (\bmod \ 2)$ given a blackbox oracle function.

Grover’s Search \cite{grover} is a well-known quantum algorithm for searching through
unstructured collections of records for particular targets. It achieves quadratic speedup compared to classical algorithms.
Given a set $X$ of $N$ elements $X=\{x_1,x_2,\ldots,x_N\}$
and a boolean function $f : X \rightarrow \{0,1\}$, the goal of an
Grover's Search is to find an element $x^* \in X$ such
that $f(x^*)=1$.
    
The Simon algorithm \cite{simon} finds a hidden integer $s \in \{0,1\}^n$ from an oracle
$f_s:\{0,1\}^n \rightarrow \{0,1\}^n$
that satisfies $f_s(x) = f_s(y)$ if and only if $y=x \oplus s$ for all
$x \in \{0,1\}^n$.

The Quantum Phase Estimation (QPE) \cite{nielsen} estimates $\theta$ such that  
$U\ket{\psi}=e^{2\pi i \theta} \ket{\psi}$,
where $U$ is a unitary operator, $\ket{\psi}$ is an eigenvector, and $e^{2\pi i \theta} $ is the corresponding eigenvalue.

The Quantum Evolution of Hamiltonian (EOH) algorithm \cite{nielsen} 
simulates the time evolution of a quantum mechanical system in which 
the system is described by a Hamiltonian $H$. 
The goal is to find an algorithm that approximates $U$  such that $||U-e^{-iHt}||\leq \epsilon$, 
where $e^{{-iHt}}$ is the ideal evolution and the maximum simulation error targeted is $\epsilon$.

The numbers of qubits simulated are 3, 5, 7 and 11 for Deutsch, Bernstein, Grover and Simon, and
3, 5, 7 for QPE and EOH. Higher numbers of qubits are not included because the algorithms always fail if an error is injected in those cases.
\begin{figure*}[ht]
\centering
     \begin{subfigure}
         \centering
\includegraphics[width=0.37\textwidth]{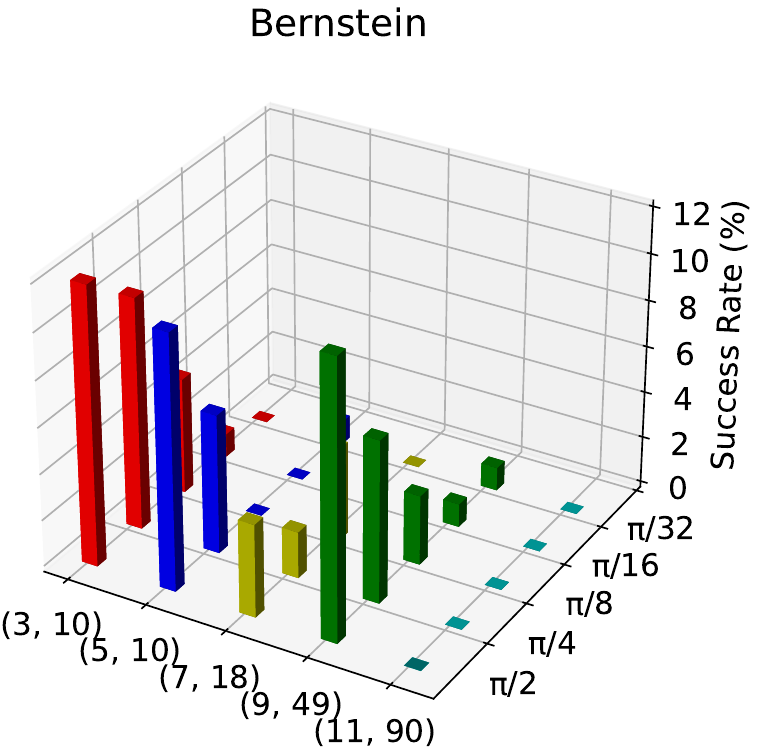} 
         \label{fig:berncoh}
     \end{subfigure}
     \begin{subfigure}
         \centering
\includegraphics[width=0.37\textwidth]{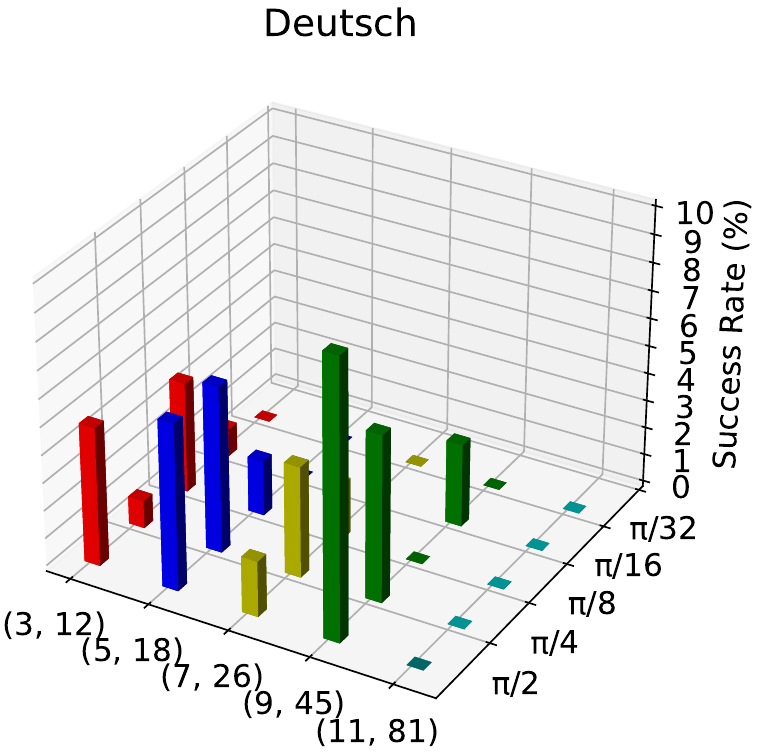}
         \label{fig:deutschcoh}
     \end{subfigure}
     \begin{subfigure}
         \centering
\includegraphics[width=0.37\textwidth]{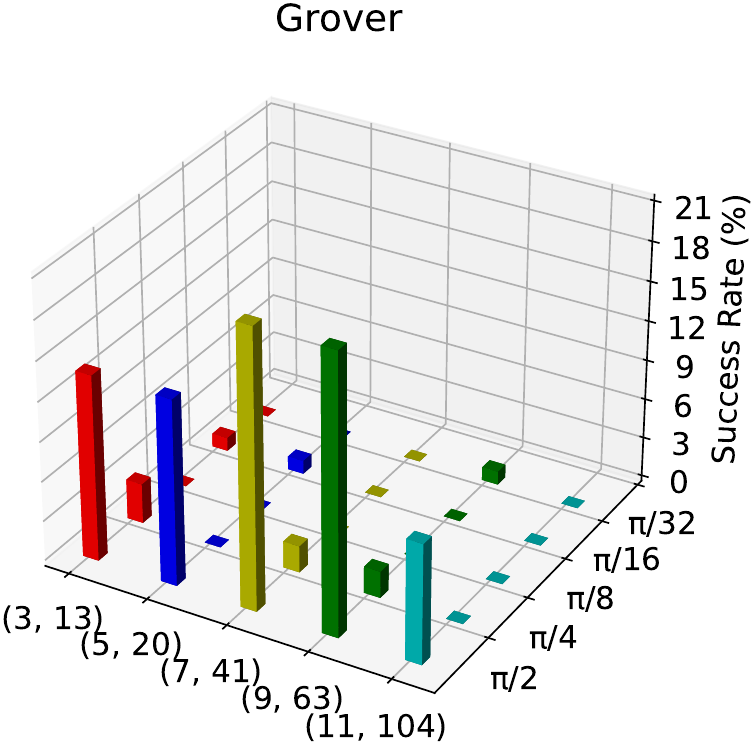}
         \label{fig:grovecoh}
     \end{subfigure}
     \begin{subfigure}
         \centering
\includegraphics[width=0.37\textwidth]{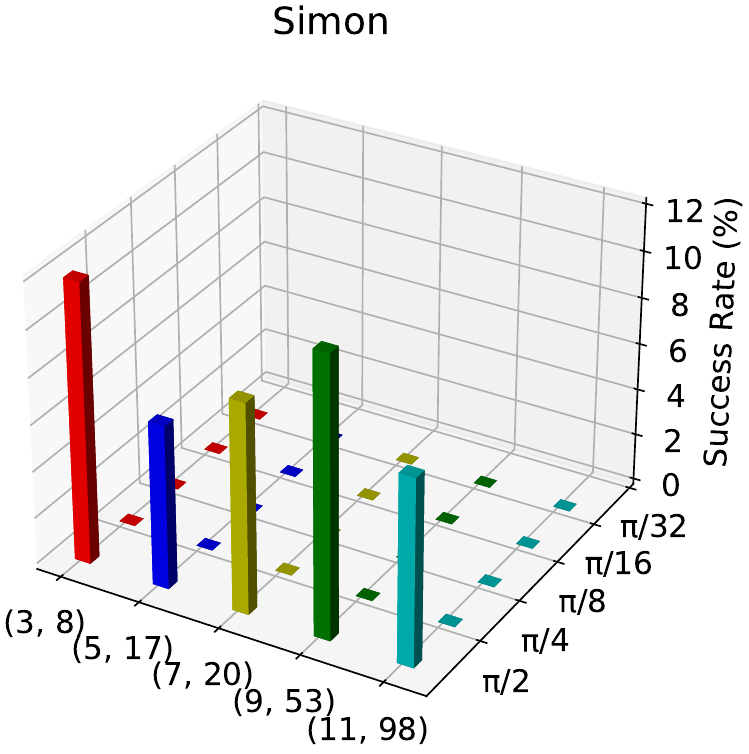}
         \label{fig:simoncoh}
     \end{subfigure}
      \caption{Single Z-rotation Error Results (\%)}
        \label{fig:zrotation}
\end{figure*}
\subsection{Results}
Figure \ref{fig:paulires} shows the success rate of the algorithms under single Pauli errors as the number of qubits increases. The x-axis  shows the number of qubits and the depth of the corresponding circuit. The y-axis is the success rate (\%).
We see that as the number of qubits and circuit depth increase, the success rate decreases. This is due to the fact that the impact of an error accumulates as the circuits get bigger with the increasing number of qubits and the circuit size. There is a single exception with Deutsch with 5 qubits where the success rate is bigger than that of 3-qubit Deutsch. 

Figure \ref{fig:zrotation} shows the success rates of the algorithms under single Z-rotation errors with different error rotation angle around Z-axis of the Bloch sphere. The x-axis  shows the number of qubits and the depth of the corresponding circuit. 
The y-axis shows the rotation angles. The z-axis is the success rate (\%).
Same with Pauli errors, under single Z-rotation errors, as the number of qubits increases, the success rates decrease. However, compared to Pauli errors, the success rates are lower showing that the resilience to the Z-rotation error is lesser.
With respect to the rotation angle, 
even though it is not monotonic, as the angle decreases, the resilience decreases too.
The decrease in an angle means that the real part of the error increases. That is why when the angle decreases so is the success rate.
We did not plot EOH and QPE because their success rates are zero in all cases.

Considering intrinsic algorithmic resilience, there are certain emerging behaviours that hold for both stochastic and coherent errors despite of the different success rates and different circuit sizes. EOH has the least relative algorithmic resilience. Following EOH, QPE and Simon have relatively less resilience to errors compared to Grover, Bernstein and Deutsch. Among these three algorithms, Bernstein and Deutsch are the most resilient.

The success rates of the algorithms under double Pauli and coherent errors are zero in all cases except for 
3 qubits. The success rates are 1\%-6\% for 3 qubits.

\section{Related Work}
\label{related}
There has been scarce work on the impact of (logical) errors on quantum algorithms. Koch et al. \cite{koch2020simulating} study the impact of gate fidelity - among other parameters -  on small circuits of Bernstein-Vazirani, Grover and Quantum Fourier Transform (QFT) algorithms.
As expected, they find the higher the fidelity the higher the success rates. They only evaluate 4-qubit circuits, which in turn limit their study. In comparison, we study higher numbers of qubits which enables us to evaluate the impact of higher number of qubits and circuit size.
Moreover, our study covers six quantum algorithms namely Bernstein-Vazirani, Grover, Simon, Deutsch-Jozsa, QPE, and EOH. As another study that investigates the effect of errors, Qi et. al. \cite{qi2022quantum} define the quantum vulnerability factor inspired by the concept of the architectural vulnerability factor in classical computing to assess the success of quantum algorithms.

Many noise-impact studies focus on certain types of quantum algorithms, such as those on variational algorithms \cite{ding2022evaluating, fontana2021evaluating, quiroz2021quantifying, sharma2020noise}.
Similarly, others focus on specific 
quantum circuits \cite{Reiner_2018, xue2019effects}.
For instance, Reiner et al. \cite{Reiner_2018} study the impact of gate errors on the time evolution of the quantum fermionic systems. As they evaluate gate errors due to over-rotations, they report that the impact of the errors depends on an algorithm's implementation.

In a different line of research, 
Willsch et al. \cite{PhysRevA.96.062302} study the dynamics of a quantum system by simulating the time-dependent Schrodinger equation. 
They evaluate the success of the gates by the metrics of the gate fidelity, the diamond distance and the unitarity. 
Interestingly, they find that the success of the gates with respect to these metrics does not reflect their performance within an algorithm. To this end, our study evaluates the success of quantum algorithms without relying on indirect gate metrics.

\section{Conclusion}
\label{conclusion}
In this work, we explore the quantum jump method and use it to evaluate and analyze 
the impact of logical stochastic Pauli and Z-rotation errors on six quantum algorithms.
The results show that as the number of qubits and the algorithm depth increase, the success rates
decrease. Moreover, they show that if two errors occur during an algorithm execution, the success rates
become zero with the exception of small-size 3-qubit algorithms. Our
results additionally show that Z-rotation errors reduce the success rates more than Pauli errors do.
With regard to algorithmic resilience, we can place the algorithms in two groups.
The algorithms of EOH, Simon and QPE are less resilient than those of Grover, Deutsch-Jozsa and Bernstein-Vazirani. Finally, our results indicate that the success rates are negatively correlated with the angle of Z-rotation errors.
 

In the future, we plan to investigate different error distributions than the uniform distribution.

\section*{Acknowledgments}
Pacific Northwest National Laboratory is operated by Battelle Memorial Institute for the U.S. Department of Energy under Contract No. DE-AC05-76RL01830. This research was supported by PNNL’s Quantum Algorithms, Software, and Architectures (QUASAR) LDRD Initiative.

\bibliographystyle{unsrt}
\bibliography{quantum_algs.bib}

\end{document}